\documentclass[a4paper,twocolumn,prb]{revtex4}
\usepackage{amsmath}
\usepackage{graphicx}
\usepackage{color}
\usepackage{subfigure}
\usepackage{hyperref}
\begin{document}
\title{Observation of SLE$(\kappa,\rho)$ on the Critical Statistical Models}
\author {M. N. Najafi }

\affiliation{Physics department, Sharif University of Technology, P.O. Box 11155-9161, Tehran, Iran}
\author{Saman Moghimi-Araghi}
\affiliation{Physics department, Sharif University of Technology, P.O. Box 11155-9161, Tehran, Iran}
\author{S. Rouhani}
\affiliation{Physics department, Sharif University of Technology, P.O. Box 11155-9161, Tehran, Iran}
\begin{abstract}
        Schramm-Loewner Evolution (SLE) is a stochastic process that helps classify critical statistical models using one real parameter $\kappa$. Numerical study of SLE often involves curves that start and end on the real axis. To reduce numerical errors in studying the critical curves which start from the real axis and end on it, we have used hydrodynamically normalized SLE($\kappa,\rho$) which is a stochastic differential equation that is hypothesized to govern  such curves. In this paper we directly verify this hypothesis and numerically apply this formalism to the domain wall curves of the Abelian Sandpile Model (ASM) ($\kappa=2$) and critical percolation ($\kappa=6$). We observe that this method is more reliable for analyzing interface loops.
\end{abstract}
\maketitle
\section{Introduction}

    Classification of probability measures on random curves in simply connected domains of the complex plane was first proposed by O. Schramm\cite{Shramm}. According to this idea one can describe the interfaces of two dimensional statistical models via growth processes named as SLE. This technique has become a powerful tool to study the macroscopic interfaces of two dimensional systems. Another powerful tool for classification of 2d critical phenomena is conformal field theory (CFT). Hence a connection between the two methods is natural.  The relation between CFT and SLE was found in the pioneering works of M. Bauer and D. Bernard \cite{BauBer}.
Parallel to the analytical achievements, a large amount of numerical works have been done using SLE to study the statistics of the critical interfaces of various statistical models. Since the interfaces of the critical models resulting from simulation or experiment are commonly loops, the customary method that is used to make curves going from origin to the infinity is to cut the loops from its middle and send the end point of the cut curves to infinity (infinity mapping). This method is widely used e.g. in Turbulence\cite{BerBofCelFal}, ASM avalanche frontier \cite{saberi1} iso height lines of KPZ  \cite{saberi2} and WO3  \cite{saberi3} , Ising model  \cite{saberi4} etc. But there is an important source of numerical error due to the infinity map which sends the end point to the infinity. This map enlarges the lattice constant in the regions close to the end point of the curve and this causes numerical large errors. To avoid these errors, we should develop a method that is free of infinity mapping.

The new method we introduce in this paper, is based on a variant of SLE$_\kappa$, known as SLE($\kappa,\rho$).  SLE($\kappa,\rho$) is a generalization of the SLE$_\kappa$ that describes the curves that are self similar but are not conformally invariant i.e. there are some preferred points on the domain that the SLE curve is growing. It has been claimed\cite{BauBer2} that the hydrodynamically normalized stochastic differential equation governing the mappings of the critical curves that start from the real axis and also end on it, is the SLE($\kappa,\rho$) where $\rho=\kappa-6$. Such maps are the ones we use to develop our numerical method. Therefore we not only introduce a new more reliable numerical method to analyze SLE curves, but also we will have a numerical check for the application of SLE($\kappa,\rho$) formalism to the curves going from the real axis to the real axis as stated above. We show that this formalism is properly applicable and obtain more precise results for the ASM and percolation. For this end, we use the slit uniformizing map to obtain the driving function and use the Maximum Likelihood Estimation (MLE) to fit properly the resulting process to the Brownian Motion to find the parameters $\rho$  and  $\kappa$ and their stochastic errors.
In section \ref{SLE} we briefly introduce chordal SLE. Sections \ref{stat} is a brief introduction to the critical percolation and ASM respectively and SLE($\kappa,\rho$). In section \ref{num} we present our numerical results.

\section{SLE}\label{SLE}
Critical behaviour of the two dimensional statistical models can be described by their geometrical features. In fact instead of studying the local observables, we can focus on the interfaces of two dimensional models. These domainwalls are some non-intersecting curves which directly reflect the status of the system in question and supposed to have two properties: conformal invariance and the domain Markov property\cite{BauBer2}. Schramm- Loewner Evolution is the candidate to analyze these random curves by classifying them to the one-parameter classes (SLE$_{\kappa}$). 
\subsection{Chordal SLE}
Let us denote the upper half plane by  $H$ and  $\gamma_{t}$ as the SLE trace i.e.  $\gamma_{t}=\lbrace z\in H:\tau_{z}\leq t \rbrace$ and the hull $K_{t}=\overline{\lbrace z\in H:\tau_{z}\leq t \rbrace}$.   $SLE_{\kappa}$ is a growth processes defined via conformal maps which are solutions of Loewner's equation:
\begin{equation}
\partial_{t}g_{t}(z)=\frac{2}{g_{t}(z)-\xi_{t}}
\label{Loewner}
\end{equation}
Where the initial condition is $g_{t}(z)=z$  and $\xi_{t}=\sqrt{\kappa}B_{t}$ is a real valued smooth function. For fixed $z$, $g_{t}(z)$ is well-defined up to time $\tau_{z}$ for which $g_{t}(z)=\xi_{t}$. The complement $H_{t}:=H\backslash{K_{t}}$ is simply connected and $g_{t}(z)$ is the unique conformal mapping $H_{t}\rightarrow{H}$ with $g_{t}(z)=z+\frac{2t}{z}+O(\frac{1}{z^{2}})$ as $z\rightarrow{\infty}$ that is known as hydrodynamical normalization. One can retrieve the SLE trace by $\gamma_{t}=\lim_{\epsilon\downarrow{0}}g_{t}^{-1}(\xi_{t}+i\epsilon)$. There are phases for these curves, $2\leq\kappa\leq{4}$ the trace is non-self-intersecting and it does not hit the real axis;  in this case the hull and the trace are identical: $K_{t}=\gamma_{t}$. This is called "dilute phase". For $4\leq\kappa\leq{8}$, the trace touches itself and the real axis so that a typical point is surely swallowed as $t\rightarrow\infty$ and $K_{t}\neq\gamma_{t}$. This phase is called "dense phase". There is a connection between the two phases: for $4\leq\kappa\leq{8}$ the frontier of $K_{t}$, i.e. the boundary of $H_{t}$ minus any portions of the real axis, is a simple curve which is locally a SLE$_{\tilde{\kappa}}$ curve with $\tilde{\kappa}=\frac{16}{\kappa}$, that is it is in dilute phase. 
The main question "what is the relation between SLE and CFT" is answered by M. Bauer and D. Bernard\cite{BauBer}. They showed that the boundary condition changing (bcc) operator in SLE correspond to a degenerate field with a vanishing descendant at level two and conformal weight $h_{1;2}=\frac{6-\kappa}{2\kappa}$ in CFT with central charge $c_{\kappa}=\frac{(6-\kappa)(3\kappa-8)}{2\kappa}$. Note that this relation is consistent with the duality expressed above: $c(\kappa)=c(\tilde{\kappa}=16/\kappa)$.

\subsection{SLE($\kappa,\rho$)}
Now we will give a brief introduction to SLE($\kappa,\rho$). We define SLE($\kappa,\rho$) in the upper half plane. The parameter $\kappa$ identifies the local properties of the model in hand and corresponds directly to its central charge, the parameter $\rho$ has to do with while the boundary conditions (bc) imposed. One bc occurs at the origin (the starting point of the curve) and the other at some point $x_{\infty}$ on the real axis. The stochastic equation governing on such curves is the same as formula (\ref{Loewner}) but the driving function has different form, namely:
\begin{equation}
d\xi_{t}=\sqrt{\kappa}dB_{t}+\frac{\rho}{\xi_{t}-g_{t}(x_{\infty})}dt
\label{SLE(k,r)}
\end{equation}
Now consider a curve that starts from origin and end on a point on real axis ($x_{\infty}$). Then by using the map $\phi={x_{\infty}z}/{(x_{\infty}-z)}$, one can send the end point of the curve to the infinity. In this respect, the function $h_{t}=\phi{o}g_{t}{o}\phi^{-1}$ describes chordal SLE. It is easy to show that the equation governing on $h_{t}$ is $\partial_{t}g_{t}=2/(\lbrace{\phi'(g_{t})(\phi(g_{t})-\xi_{t})}\rbrace)$. But it is explicit that this function is not hydrodynamically normalized. It has been shown\cite{BauBer2} that if one uses another mapping $\tilde{g}_{t}=v_{t}{o}h_{t}{o}u^{-1}$ where $u=\phi^{-1}$ and $v_{t}$ is any linear fractional transformation that make the corresponding map hydrodynamically normalized, then the stochastic equation of $\tilde{g}_{t}$ is the same as Eq. (\ref{Loewner}). In fact this procedure leaves the Eq. (\ref{Loewner}) unchanged but leads the driving function to have a drift term\cite{BauBer2}:
\begin{equation}
d\xi_{t}=\sqrt{\kappa}dB_{t}+\frac{\kappa-6}{\xi_{t}-g_{t}(x_{\infty})}dt
\label{driving}
\end{equation}
In the other words, this stochastic function is the driving function of the SLE($\kappa,\rho$) with $\rho=\kappa-6$. Thus for the critical curves from boundary to boundary, the corresponding driving function acquires a drift term. This generalization of SLE can be generalized further to have multiple preferred real axis points. For review see[13].
\section{statistical models}\label{stat}
In this section we briefly introduce some statistical models to be simulated in the next section. To cover the argument properly we choose models from two regimes of $\kappa$ that are ASM with $\kappa=2<4$ (dilute phase) and percolation with $4<\kappa=6<8$ (dense phase). In below we give a short review of the two models.
\subsection{Critical Percolation}
\begin{figure}
\centerline{\includegraphics[scale=.40]{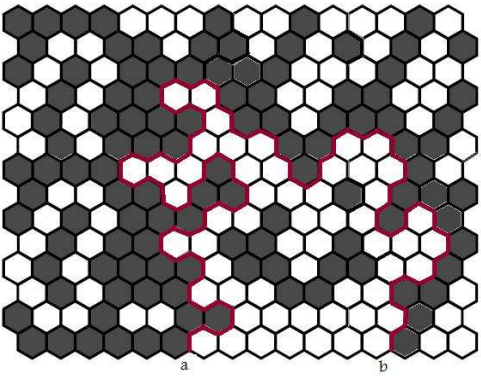}}
\caption{A typical configuration of the percolation starting and ending on the real axis.}
\label{DASM}
\end{figure}

Let (H, a, b) be an hexagonal lattice domain on the upper half plane with the boundary condition changes on two points on the real axis namely a and b i.e. color the hexagons on the boundary in black and white so that the resulting boundaries have two boundary condition changes on points a and b as indicated in the Fig. \ref{DASM} . A configuration is a choice of color for inner hexagons which becomes black or white with the probability $p$ and $1-p$ respectively. One can easily identify unique domainwalls separating white and black sites from each other. Each configuration defines an interface, i.e. the unique path from a to b in H such that the hexagon on the left (resp. right) of any of its edges is black (resp. white), see Fig. \ref{DASM}. Hence the probability distribution on configurations induces a probability distribution on paths from a to b in H. There is a critical probability $p=p_{c}=0.5$ in which the size of the resulting clusters is of order of the size of lattice and the clusters are self similar. We consider here the critical interfaces of percolation which is known that are SLE($\kappa=6$).
Percolation enjoys also an important property namely locality, that is, the evolution of the SLE curve is insensitive to the boundary conditions and the events that take place in the $H\setminus{K_{t}}$ up to the time $t$. Due to the locality, the interfaces of the percolation can be generated by a simple growth process: in the n'th step, when the tip of the curve reaches to an inner hexagon, then using a fair coin, color this site in black or white. The upshot is that the $n+1$'th step of the resulting domainwall will be the edge of the hexagon whose adjacent faces have different colors. By this simple rule, we can generate domainwall samples that start from the origin and end on the arbitrary point on the lattice boundary. Locality implies that this growth process is insensitive to the fact that "where the curve will end". It is guaranteed in Eq. (\ref{driving}) which implies $\rho=\kappa-6=0$, saying that the drift term in this equation vanishes. So these stochastic curves evolve like the chordal one and are insensitive to the boundary conditions as expected.
 \subsection{ASM} 
Sandpile models have been introduced by Bak, Tang and Wiesenfeld\cite{Bak} as an example for a class of models that show Self-Organized Criticality(SOC). These models show critical behaviours, without tuning external parameters such as temperature. The abelian structure of this model was first discovered by D. Dhar and named as Abelian Sandpile Model (ASM)\cite{Dhar1}. Despite its simplicity, ASM has various and interesting features and many different analytical and numerical works have been done on this model. For example different height and cluster probabilities\cite{Majumdar3}, its connection with spanning trees\cite{Majumdar2}, ghost models\cite{Mahieu}, q-state potts model\cite{Saluer}, etc. For a good review see [16].

Consider the ASM on a two dimensional square lattice $L\times{L}$. To each site $i$, a height variable $h_{i}$ is assigned taking its values from the set {1,2,3,4} the number of sands in this site. The dynamics of this model is as follows; in each step, a grain is added to a random site $i$ i.e. $h_{i}\rightarrow{h_{i}+1}$; if the resulting height becomes more than 4, the site toppels and loses 4 sands, each of which is transfered to one of four neighbors of the original site. As a result, the neighboring sites may become unstable and topple and a chain of toppling may happen in the system. In the boundary sites, the toppling causes to one or two sands to leave the system. This process continues until the system reaches to a stable configuration. Now another random site is selected and the sand is released on this site and the process continues. The movement on the space of stable configuration lead the system to fall in a subset of sets of configurations after a finite steps, named as the "recurrent states". For details see[16]. This model is related to Potts model with $q=0$ and CFT with $c=0$. For a lattice with $d$ neighboring sites, the toppeling occurs when $h_{i}>d$, then the original site will lose $d$ grains and the height of each of its neighbors will increase by 1. It is obvious that for a triangular lattice $d=6$.

\section{Numerical Methods and Results}\label{num}
\begin{figure}
\centerline{\includegraphics[scale=.50]{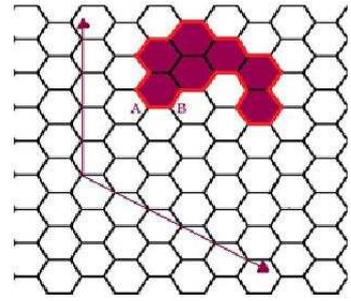}}
\caption{The schematic view of triangular lattice with two sublattices A and B.}
\label{AB lattice}
\end{figure}

To use Eq. (\ref{Loewner}) for the statistical models on the lattice, one has to discretize this equation. For this purpose it is customary to assume an especial function form for the driving function in each discrete interval and find the corresponding uniformizing map in that interval and using $G_{t_{n}}=G_{t_{n-1}}{o}G_{t_{n-2}}{o}...{o}G_{t_{1}}$ send every point on the $\gamma_{t}$ to the real axis step by step. If we assume that $\xi$ is constant in an interval ($\xi_{n}$), the corresponding uniformizing map is
\begin{equation}
G_{t_{n}}(z)=\xi_{n}+\sqrt{(z-\xi_{n})^{2}+4t_{n}}
\label{slit}
\end{equation}                                                                                               
It is notable that $g_{t}(z)=z+\frac{2t}{z}+O(\frac{1}{z^{2}})$ as $z\rightarrow{\infty}$. The method that is widely used to apply this formalism to the interface loops for the statistical models on lattice is as follows: one cuts the loops horizontally from its middle to generate curves starting from the origin and ending at some point on the real axis. Then by mapping the end point of the curve to infinity and applying the chordal SLE i.e. Eq. (\ref{Loewner}) one can obtain the underlying driving function.
 For the simulation, we consider the triangular lattice and its dual lattice (honeycomb lattice) and also consider two sublattices A and B as indicated in Fig. \ref{AB lattice}. Thus the position of each point can be coded in ($n,m$,A/B) where $n$ and $m$ shows the position of the lattice point and A or B shows its basis. It is obvious that on this lattice, each point has 6 neighbours.
Consider that we have ensembles of interface loops of some model, then by the process described above we have ensembles of the curves that start and end on the real axis. Thus when we have the set of discrete points $\lbrace{z_{1},z_{2},z_{3},...,z_{n}}\rbrace$, the first unformizing map is of the form of Eq. (\ref{slit}) with the parameters $\delta{t_{1}}=\frac{1}{4}(\text{Im}[z_{1}])^{2}$ and $\xi_{1}=\text{Re}[z_{1}]$ and after applying this mapping we continue this process for $z_{2}$. The process continues until every point of the curve is sent to the real axis. To analyze the resulting driving function and use Eq. (\ref{slit}), we discretize this equation.
\begin{equation}
\delta\xi_{n}=\sqrt{\kappa}\delta{B_{n}}+\frac{\rho}{\xi_{n}-G_{t_{n}}(x_{\infty})}\delta{t_{n}}
\label{dis driving}
\end{equation}
For having a pure Brownian motion, we should write the Eq. (6) of the form;
\begin{equation}
\frac{\xi_{n}-\sum_{i=1}^{n}[\frac{\rho\delta{t_{i}}}{\xi_{i}-G_{t_{i}}(x_{\infty})}]}{\sqrt{\kappa}}=B_{n}
\label{BM}
\end{equation}
Where $\rho=\kappa-6$. After doing this process, we can analyze the resulting quantity on the right hand side of Eq. (\ref{BM}) to see if this stochastic process is Brownian Motion. The best fit to the Brownian Motion, gives us the amount of $\kappa$. For this, we used the Maximum Likelihood Estimation (MLE) method. In this method, the best value for a model parameters (here $\kappa$ and $\rho_{0}\equiv -\rho$) are which minimize the function and their precisions can be obtained from the distribution function of the $\chi^{2}$ function by calculating the ratio of the area under the resulting diagram to the whole area under it.

\subsection{Critical Percolation}
\begin{figure}
\centerline{\includegraphics[scale=.50]{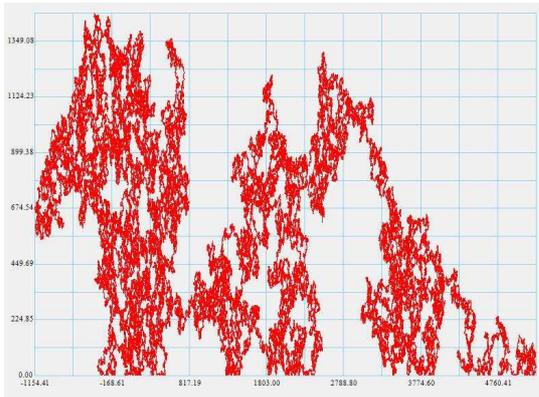}}
\caption{A sample graph of the percolation going from real axis to itself.}
\label{persam}
\end{figure}

\begin{figure}
\centerline{\includegraphics[scale=.25]{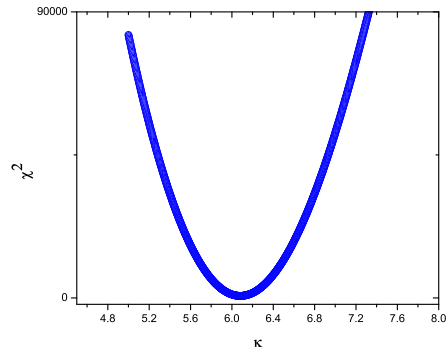}}
\caption{$\chi^{2}$ graph vs $\kappa$, it has a minimum at $\kappa=6.08$ and $\rho_{0}=-0.08$.}
\label{perchisq2}
\end{figure}

Consider the interfaces of critical percolation. A typical critical percolation curve is shown in Fig. \ref{persam} that starts from origin and ends at the point 5200 on the real axis. Due to its locality property, percolation is so hard to be controlled to end on a certain point, in this figure it reaches the real axis after 850000 steps. We simulated over 10000 curves with the typical length 50000. After simulating such curves we obtained $\xi_{t}$ as described above. The Fig. \ref{perchisq2} shows the behaviour and the minimum of the $\chi^{2}$ function versus $\kappa$ which is obtained by fitting $\langle{B_{t}^{2}}\rangle$ versus $t$. The global minimum of this curve is in the point $\kappa=6.08$ and $\rho_{0}=6-\kappa=-0.08$ in agreement with the expected values ($\kappa=6$ and $\rho_{0}=0$). This result justifies the locality property of the percolation as stated above.

\begin{figure}
\centerline{\includegraphics[scale=.35]{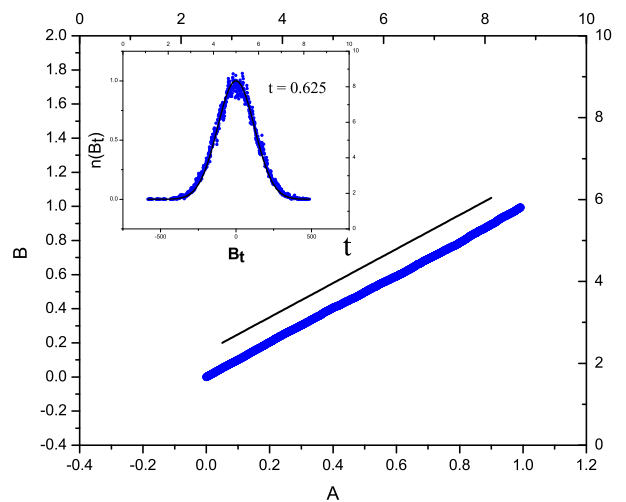}}
\caption{The graph of the obtained process $\langle B_{t}^{2}\rangle$ versus $t$ for percolation.}
\label{perB2}
\end{figure}

Using the distribution of $\chi^{2}$ we obtain the precision of the resulting parameters that is, with the probability 60\% $\kappa=6.08\pm{0.07}$ and $\rho_{0}=-0.08\pm{0.07}$. The plot of $\langle{B_{t}^{2}}\rangle$ versus $t$ graph for the obtained $\kappa$ and $\rho_{0}$ is indicated in Fig. \ref{perB2}. We see that this graph is linear with the slope $0.98\pm{0.02}$. The inner graph shows the distribution of $B_{t}$'s for $t=0.625$ that is properly fitted to the Gaussian function with $\sigma=0.625$ this ensures us that this process is also a Brownian Motion.

\subsection{ASM}
\begin{figure}
\centerline{\includegraphics[scale=.50]{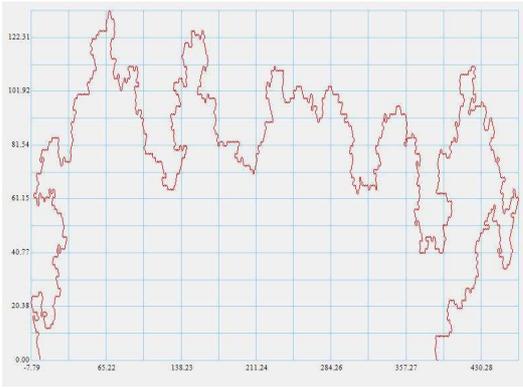}}
\caption{A sample interface of ASM starting from origin to the real axis.}
\label{ASMsam}
\end{figure}

\begin{figure}
\centerline{\includegraphics[scale=.35]{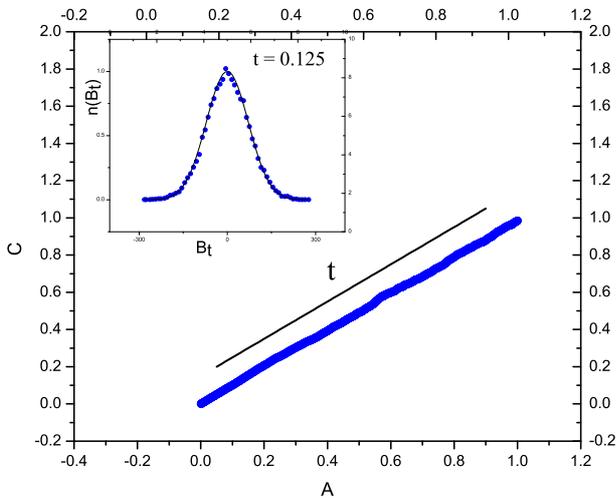}}
\caption{The $\langle B_{t}^{2}\rangle$ graph vs $t$ that is shown is linear with a good slope $1.00\mp{0.05}$ for ASM. The inner graph is the distribution of $B_{t}$  for $t=0.125$. The red graph is the numerical result and the blue one is the Gaussian distribution $\exp[-\frac{B^{2}}{2\sigma}]$ with $\sigma=0.125$.}
\label{ASMB2}
\end{figure}

\begin{figure}
\centerline{\includegraphics[scale=.30]{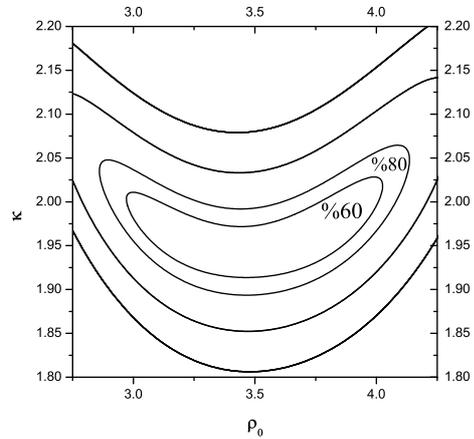}}
\caption{The contour graph for the $\chi^{2}$ (for fitting $\langle{B_{t}^{2}}\rangle$ with $t$) as a function of $\kappa$ and $\rho_{0}$.}
\label{ASMcont}
\end{figure}

Consider now avalanche frontiers of ASM. By adding a grain to the saturated sandpile and making it unstable, avalanches occur and their frontier form a loops with discrete points. Fig. \ref{ASMsam} shows a typical interface of the ASM that goes from real axis to the real axis.
As stated above from the theory one expects that $\rho_{0}=6-\kappa=4$. To test this, we let $\rho_{0}$ to be a free parameter in this case. The resulting amounts for the ASM on the 2048$\times$2048 triangular lattice over 10000 interface curves are $\kappa=1.95\pm{0.05}$ and $\rho_{0}=3.5\pm{0.5}$ with the probability 60\%. Fig. \ref{ASMcont} indicates the the contour plot of $\chi^{2}$ in terms of $\kappa$ and $\rho_{0}$ in which the probability of finding parameter values in various regions are indicated. In Fig. \ref{ASMB2} we have shown the graph $\langle{B_{t}^{2}}\rangle$ versus $t$ for the parameters considered above ($\kappa=1.95$ and $\rho_{0}=3.80$). It is seen that the graph is properly linear with the slope $0.98\pm{0.05}$ ($\langle{B_{t}^{2}}\rangle\simeq{t}$) that is the famous property of the Brownian motion. In the inner graph of Fig. \ref{ASMB2} the distribution of $B_{t}$ at time $t=0.125$ and also the Gaussian distribution $\exp[-\frac{B^{2}}{2\sigma}]$ with $\sigma=0.125$ are indicated in the same time. 

As is seen in the graph they do suitably fit to each other showing that the distribution of $B_{t}$'s are Gaussian. These evidences and the fact that $\langle{B_{t}}\rangle\simeq{0}$ show that this stochastic process is Brownian Motion and the driving function obtained from hydrodynamically normalized uniformizing map process for this process has the form of Eq. (\ref{driving}).

\section{Conclusion}
In this paper, we analysed the statistics of the curves that go from real axis to itself using the formalism of hydrodynamically normalized SLE($\kappa,\rho$) which is expected to describe them on the upper half plane, to test its validity and make the numerical calculations more exact. For this end we considered avalanche frontiers of the critical ASM and the critical percolation. Then fitting to the Brownian Motion and using the Maximum Likelihood Estimation, we calculated the best values of parameters i.e. corresponding $\kappa$ and $\rho_{0}=-\rho$ and their precisions. In this way we introduced a new method to analyze such a curves that is more precise and more reliable.

\end{document}